\documentclass[12pt]{iopart}
\usepackage{iopams} 
\usepackage{graphicx}
\usepackage{amssymb}
\usepackage{amsfonts}

\newcommand{\be}{\begin{equation}}
\newcommand{\bd}{\begin{displaymath}}
\newcommand{\ee}{\end{equation}}
\newcommand{\ed}{\end{displaymath}}
\newcommand{\ba}{\begin{eqnarray}}

\newcommand{\ea}{\end{eqnarray}}

\newcommand{\half}{{\textstyle \frac{1}{2}}}
\newcommand{\halfs}{{\scriptstyle \frac{1}{2}}}
\newfont{\mycal}{eufb10 at 12pt}
\newfont{\myeu}{eufm10}

\renewcommand{\theequation}{\arabic{section}.\arabic{equation}}
 \renewcommand{\theequation}{\arabic{equation}}

   \newcommand{\rd}{{\rm d}}
   \newcommand{\re}{{\rm e}}

\begin{document}
\title[ Ising Models with Holes]%
{Ising Models with Holes: Crossover Behavior}

\author{Helen Au-Yang and Jacques H.H. Perk}
\address{Department of Physics, Oklahoma State University, 
145 Physical Sciences, Stillwater, OK 74078-3072, USA}
\ead{helenperk@yahoo.com, perk@okstate.edu}

\begin{abstract}
In order to investigate the effects of connectivity and proximity in the specific heat, a special class of exactly solvable planar layered Ising models has been studied in the thermodynamic limit. The Ising models consist of repeated uniform horizontal strips of width $m$ connected by sequences of vertical strings of length $n$ mutually separated by distance $N$, with $N=1,2$ and $3$. We find that the critical temperature $T_{\mathrm c}(N,m,n)$, arising from the collective effects, decreases as $n$ and $N$ increase, and increases as $m$ increases, as it should be. The amplitude $A(N,m,n)$ of the logarithmic divergence at the bulk critical temperature $T_{\mathrm c}(N,m,n)$ becomes smaller as $n$ and $m$ increase. A rounded peak, with size of order $\ln m$ and signifying the one-dimensional behavior of strips of finite width $m$, appears when $n$ is large enough. The appearance of these rounded peaks does not depend on $m$ as much, but depends rather more on $N$ and $n$, which is rather perplexing. Moreover, for fixed $m$ and $n$, the specific heats are not much different for different $N$. This is a most surprising result. For $N=1$, the spin-spin correlation in the center row of each strip can be written as a Toeplitz determinant with a generating function which is much more complicated than in Onsager's Ising model. The spontaneous magnetization in that row can be calculated numerically and the spin-spin correlation is shown to have two-dimensional Ising behavior. 
\end{abstract}
\maketitle
\section {Introduction}

To gain more theoretical insight into proximity effects \cite{GKMD,KMG,PKMG,PKMGn,PG,PKMGpr,TPG} and crossover behavior, we study here the specific heats of special planar Ising models, which consist of periodically repeated strips of width $m$ lattice spacings and in which the coupling energy between the nearest-neighbor Ising spins is $J$. The strips are connected to one another by sequences of strings of length $n$ on which the pair interaction is also $J$. These strings are separated from one another by a distance $N$. This is illustrated in Fig.~1.
\begin{figure}[hbt]\begin{center}
\includegraphics[width=0.5\hsize]{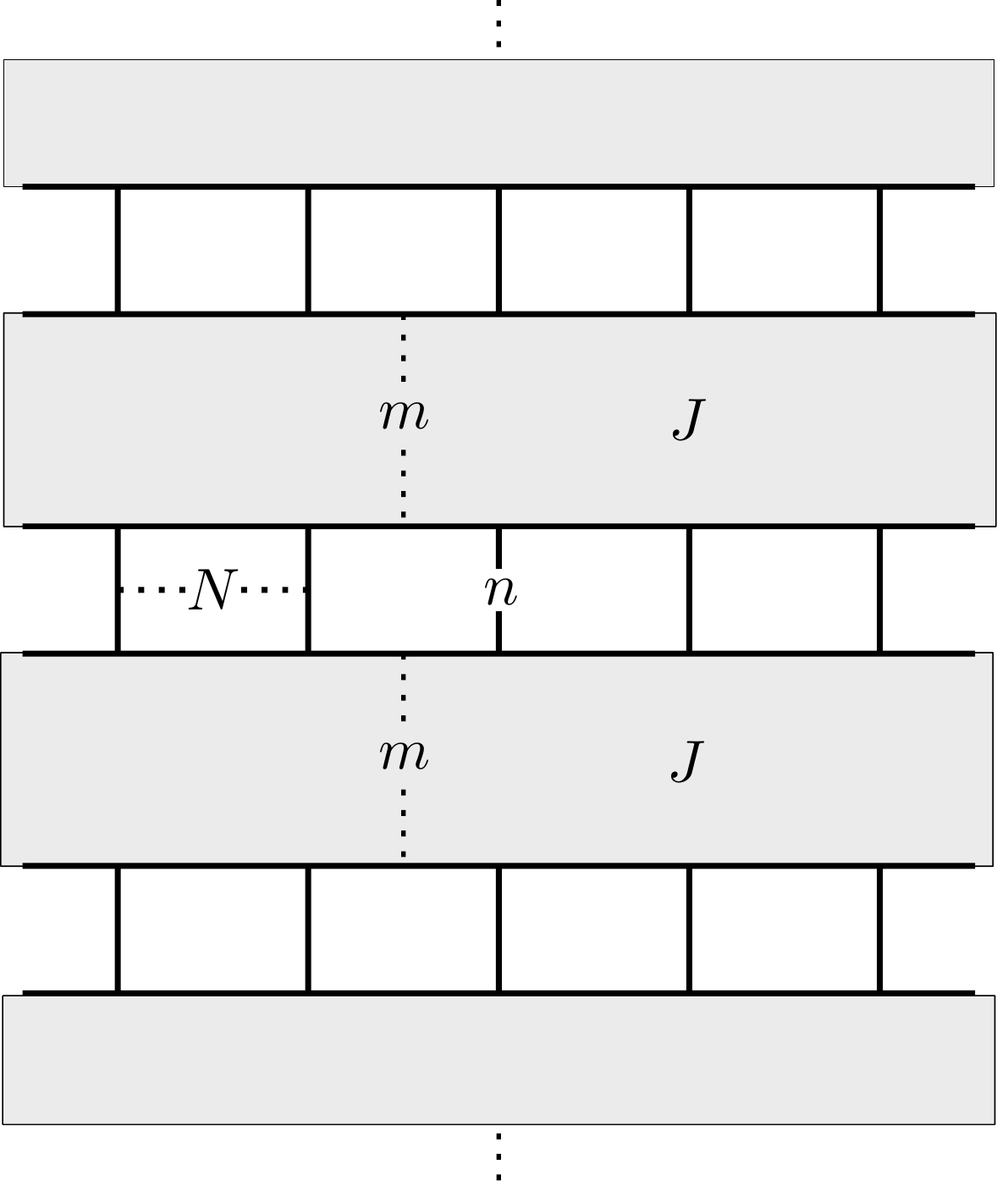}
\caption{The planar square-lattice Ising model studied consists of periodically repeated strips of width $m$ connected by sequences of strings of length $n$ separated by distance $N$. The widths $m$, $n$ and $N$ are measured in nearest-neighbor lattice spacings. There are Ising spins $\sigma_i=\pm1$ on all sites of the shaded regimes and on the strings; these spins are coupled via nearest-neighbor pair interaction $J\sigma_i\sigma_j$. Inside the white spaces there are no spins.}
\end{center}
\label{fig:1}
\end{figure} 
This model is more simple than the one considered by Abraham et al.\ \cite{AMSV}. However, due to the relative simplicity, exact calculations can be done to understand the impact of these lines connecting the strips. In the experiment reported in \cite{PKMGn}, many little boxes filled with superconducting helium are linked by thin channels which are not superconducting. These authors found that these boxes do not behave as quantum dots, but rather exhibit effects caused by being connected through these thin channels, calling this `proximity' effects. The theoretical understanding of such behavior is still inadequate. Here we use exact calculations on the Ising model to gain some further insight.

As known from the decoration method \cite{Syozibk}, a string of $n$ spins, interacting with nearest-neighbor coupling strength $J$ can be transformed to a single pair with interaction strength $ \bar J$ between the two spins located at the end of the string, satisfying the relation $ \bar z=\tanh \bar J/k_{\mathrm B}T=z^n$, where $z= \tanh J/k_{\mathrm B}T<1$. This means $\bar z=z^n\to0$ in the limit $n\to\infty$. In other words, when the length $n$ of the strings become infinite, the system behaves as a set of independent strips of width $m$. In this limiting case the system has one-dimensional behavior, so that the specific heat is not divergent, but has a rounded peak; also its spontaneous magnetization is identically zero for $T>0$. As the strip width $m$ increases, the peak of the specific heat increases and its location moves toward the original Onsager critical temperature. However, the spontaneous magnetization remains identically zero for strips of width $m$.

When the length of the strings $n$ is finite, there is a true critical temperature $T_{\mathrm c}(N,m,n)$, which is a function of the separation $N$ between the strings, the length $n$ of the strings and the width $m$ of strips. Below this critical temperature the spontaneous magnetization is non-zero. The specific heat diverges logarithmically at $T_{\mathrm c}(N,m,n)$. Thus the system behaves as the two-dimensional Ising system.

In the alternating layered Ising models studied recently in \cite{HAYFisher, HAY}, there are three critical temperatures to consider, namely, the true critical temperature at which the specific heat diverges, and the two critical temperatures of the two infinite strips of different couplings. However, here in this model we only need to consider two---the true bulk critical temperature and the critical temperature of the original Onsager Ising model at
\be
 \frac{k_{\mathrm B}T_{\mathrm c}}J=\frac2{\ln(1+\sqrt 2)}= 2.2691853142\cdots,\quad z_c=\tanh\frac{J}{k_{\mathrm B}T}_c=\sqrt{2}-1.
 \label{OnsagerTc}\ee
Consequently, we believe that this model is somewhat closer to the experiments of Gasparini et al.\ \cite{GKMD,PKMGpr}.

For $m$ and $n$ large, the specific heat has a clear rounded peak, which moves closer and closer to Onsager's critical temperature (\ref{OnsagerTc}) as $m$ increases, and the amplitude of the logarithmic divergence at the true bulk critical temperature $T_{\mathrm c}(N,m,n)$ becomes exponentially small; nevertheless it is there. For $T<T_{\mathrm c}(N,m,n)$, the spontaneous magnetization should be nonzero. This then is a model that can demonstrate the proximity effect of how one system impacts the other and to understand the possible crossover of two-dimensional to one-dimensional behavior.

To calculate the specific heat, we have used the dimer method given in \cite{MWbk}, which relates the free energy to a Pfaffian, whose square is a determinant of a sparse matrix. More specifically, we use the iteration method given on pages 120--121 in \cite {MWbk} to calculate this determinant. Such procedures were also used in early studies of layered Ising models \cite{HAYMcCoy,Hamm}.\footnote{For explicit results for the free energy of the layered Ising model with periods 3 and 4
see \cite{KB,LH}.}\ The calculation becomes very messy as $N$ increases; however, major cancellations take place making the final result not so bad. To make sure the result is correct, and for comparison too, we have also used the method described in \cite{HAYFisherFerd}, in which they calculated the determinant of a matrix $U$ by taking from matrix $U_0$ of the perfect Ising lattice, as shown in (2.21) of \cite{HAYFisherFerd}. As the difference of the two matrices is not so big, we were able to reduce the calculation of the original determinant of size $4mN\times 4mN$ to the calculation of a $2N\times 2N$ determinant. We found identical results from the two different ways, and also found that this second method is simpler. These calculations will be presented elsewhere.

For $N=1$, we have calculated the generating function of the row spin correlation function in the central row of one strip of width $m$. We thus calculated the spontaneous magnetization and the correlation functions for different values of $m$ and $n$, to understand the impact of the strings on the spins. For this we have used the Gamma matrix approach introduced by Kaufman \cite{Kaufman, KHuang} as also applied to the calculation of the correlation functions in the special row of the Bariev model \cite{McCoyPerk}. 
\section {Qualitative Observations}
\subsection {True Critical Temperature}
When $N$ or $n$ increases, the white area shown in Fig.~1, in which there are no spins (or the spins do not interact with one another), increases, and therefore the bulk critical $T_{\mathrm c}(N,m,n)$ should decrease as shown in Fig.~2(a). In the limit $n\to\infty$, the system becomes an infinite set of strips of finite width $m$, which are essentially one-dimensional Ising models. Thus the system shows no divergence in the specific heat, and zero spontaneous magnetization. On the other hand, as the strip width $m$ increases, so that the relative number of interacting spins increases, the true critical temperature $T_{\mathrm c}(N,m,n)$ increases. This is shown in Fig.~2(b). In the limit $m\to\infty$, it is two dimensional even for $n\to\infty$. Therefore it is a simplest system for which exact calculations can be done to understand the cross-over and proximity behavior.
\begin{figure}[htb] 
\centering
  \vspace*{0pt}
     \includegraphics[width=0.45\hsize]{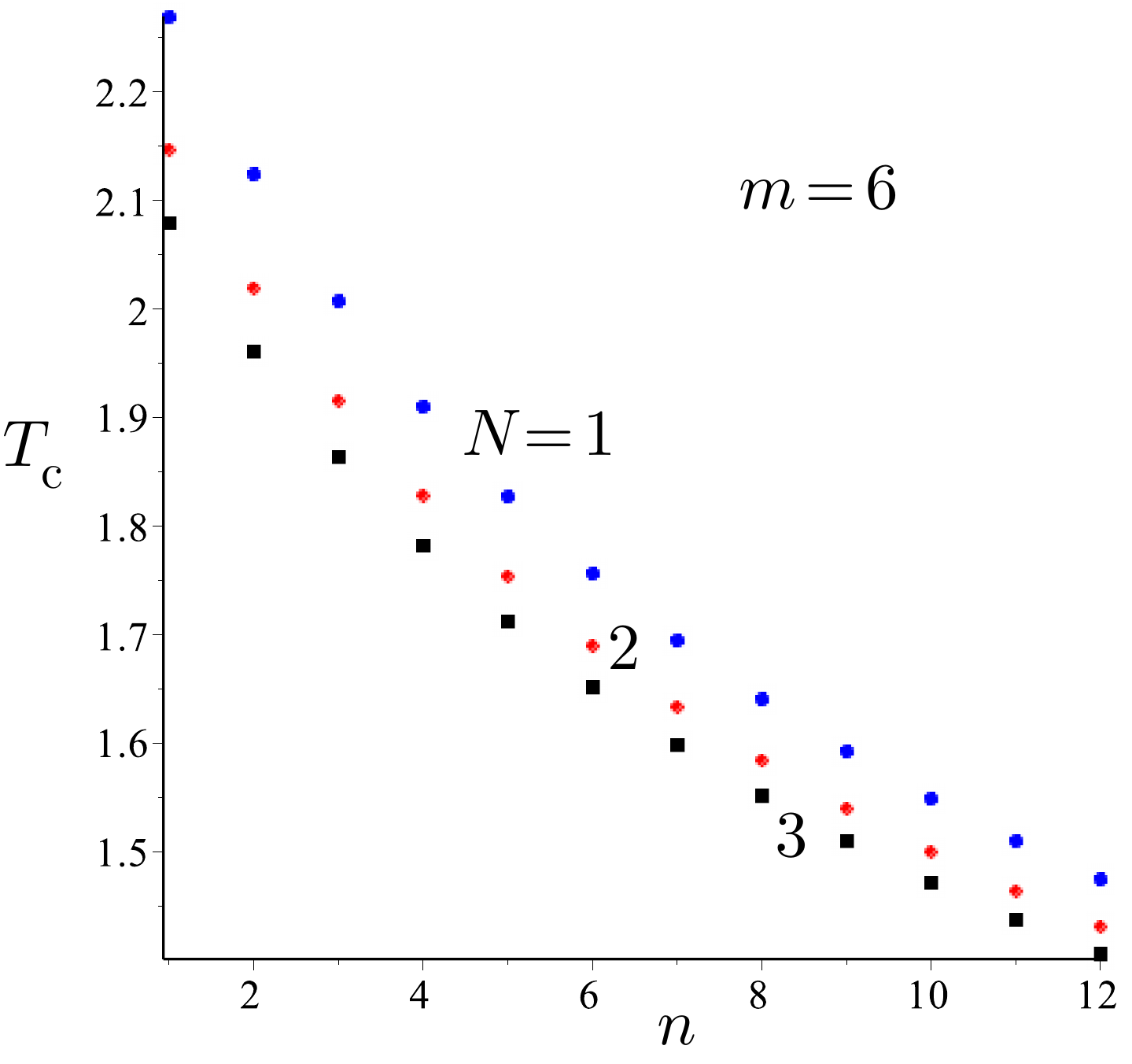}
        \includegraphics[width=0.45\hsize]{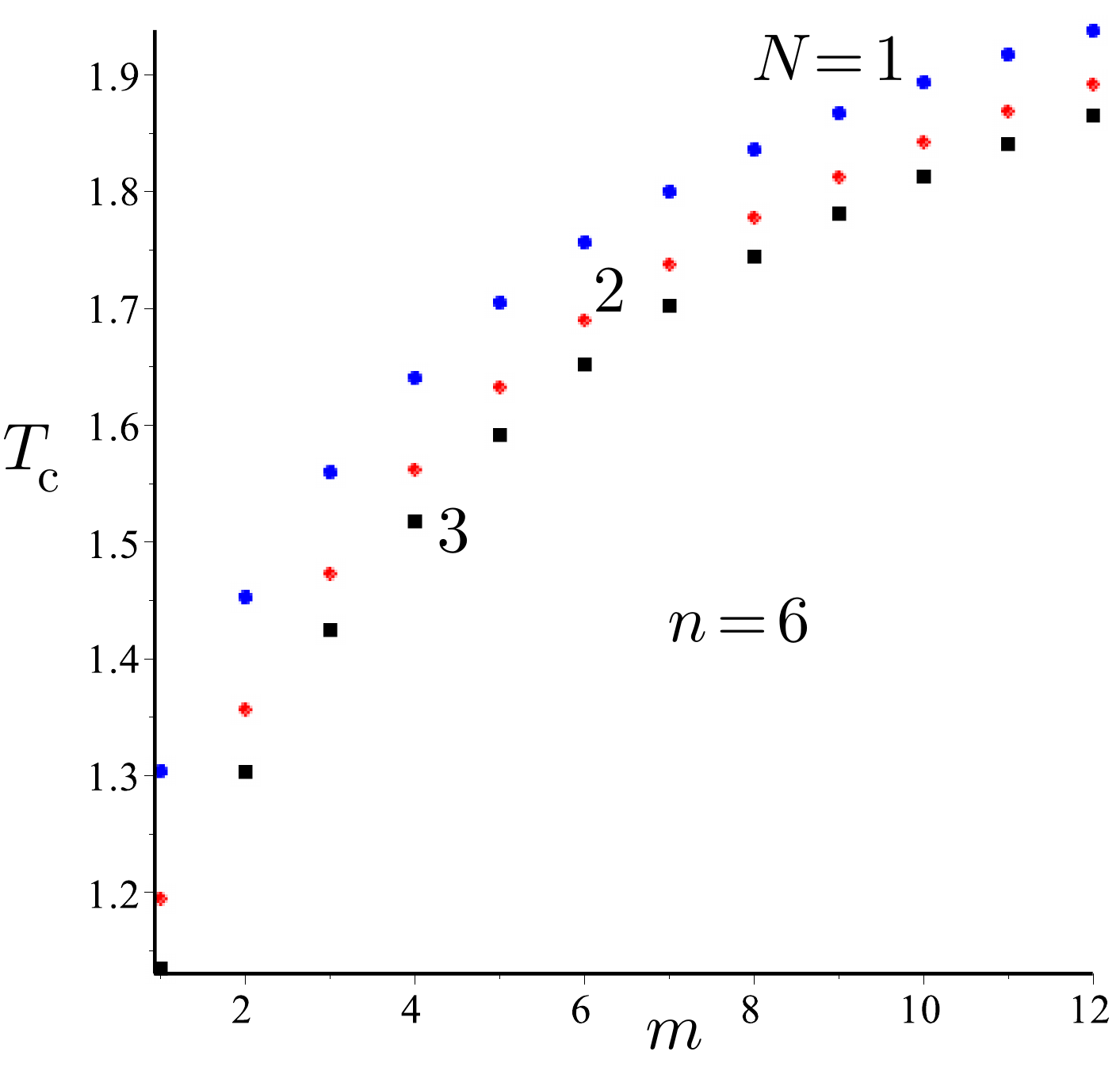} 
\caption{(Color online) (a) The critical temperature $T_{\mathrm c}(N,m,n)$ for $N=1,2,3$, is plotted for different $n$ at fixed strip width $m=6$. The points for $N=1$ denoted by blue solid circles are are above the solid diamonds for $N=2$, which in turn are above the solid squares for $N=3$. The critical temperature $T_{\mathrm c}(N,m,n)$ decreases as $n$ increases. %
(b) Plots of $T_{\mathrm c}(N,m,n)$ for fixed $n=6$, and $N=1,2,3$, as functions of $m$. Now $T_{\mathrm c}$ increases as $m$ increases. }
\label{fig:2}
\end{figure}
\subsection {N=1}
For $N=1$, the strings are next to each other, and the calculation of the free energy is straightforward and simple. It is also a special case of the layered model considered earlier \cite{HAYFisher, HAY}. The critical temperature is determined by
\be(1-z)/(z+1)=z^{(m+n)/(m+1)},\quad z=\tanh (J/k_{\mathrm B}T).
\label{criticalT1}\ee
For $n=1$, it is Onsager's Ising model, and its specific heat diverges at $z_c=\sqrt 2-1$, as also seen from (\ref{criticalT1}). As $n$ increases, the bulk critical temperature $T_{\mathrm c}(1,m,n)$ determined from this equation becomes smaller. We find that, for $1<n\le 4$, the specific heat diverges at $T_{\mathrm c}(1,m,n)$ logarithmically in the same manner as in the regular Ising model. The specific heat as a function of temperature is plotted in Fig.~3(a), for $n=4$, $m=4, 6,12$, showing divergent behavior for all $m$. The critical temperatures $T_{\mathrm c}(1,m,4)$ denoted by dashed vertical lines are now lower than Onsager's critical temperature given in (\ref{OnsagerTc}) and denoted by a solid vertical line, but they move toward it as $m$ increases. A rounded peak does not appear for any $m$, and this is true for $n\le 4$. 
In Fig.~3b, we plot the specific heat per site as a function of temperature for fixed width of strip $m=12$ and for various values of $n$. We show that for $n=5$, a small rounded peak appears just above the bulk critical temperature $T_{\mathrm c}(1,12,5)$. As $n$ increases, this rounded peak (due to the 2d strips) becomes more prominent, while the amplitude of the logarithmic divergence at the ``true" bulk critical point $T_{\mathrm c}(1,m,n)$, which dominates for $n\le 5$, decreases rapidly as $n$ increases, and this singularity becomes almost invisible for $n$ large. When the strings become longer another peak shows up at low temperature (due to these strings), corresponding to the maximum of the specific heat of the 1D Ising model at $T_{\rm max}$,
\be
(J/k_{\mathrm B}T_{\rm max})\tanh(J/k_{\mathrm B}T_{\rm max})=1,\quad k_{\mathrm B}T_{\rm max}/J=0.8335565596\cdots.
\ee
This is seen in the plot for $n=11$.
For $n\ne\infty$, we shall show in the next section that the spontaneous magnetization is nonzero for $T<T_{\mathrm c}(1,m,n)$.
\begin{figure}[htb] 
\centering
\includegraphics[width=0.48\hsize]{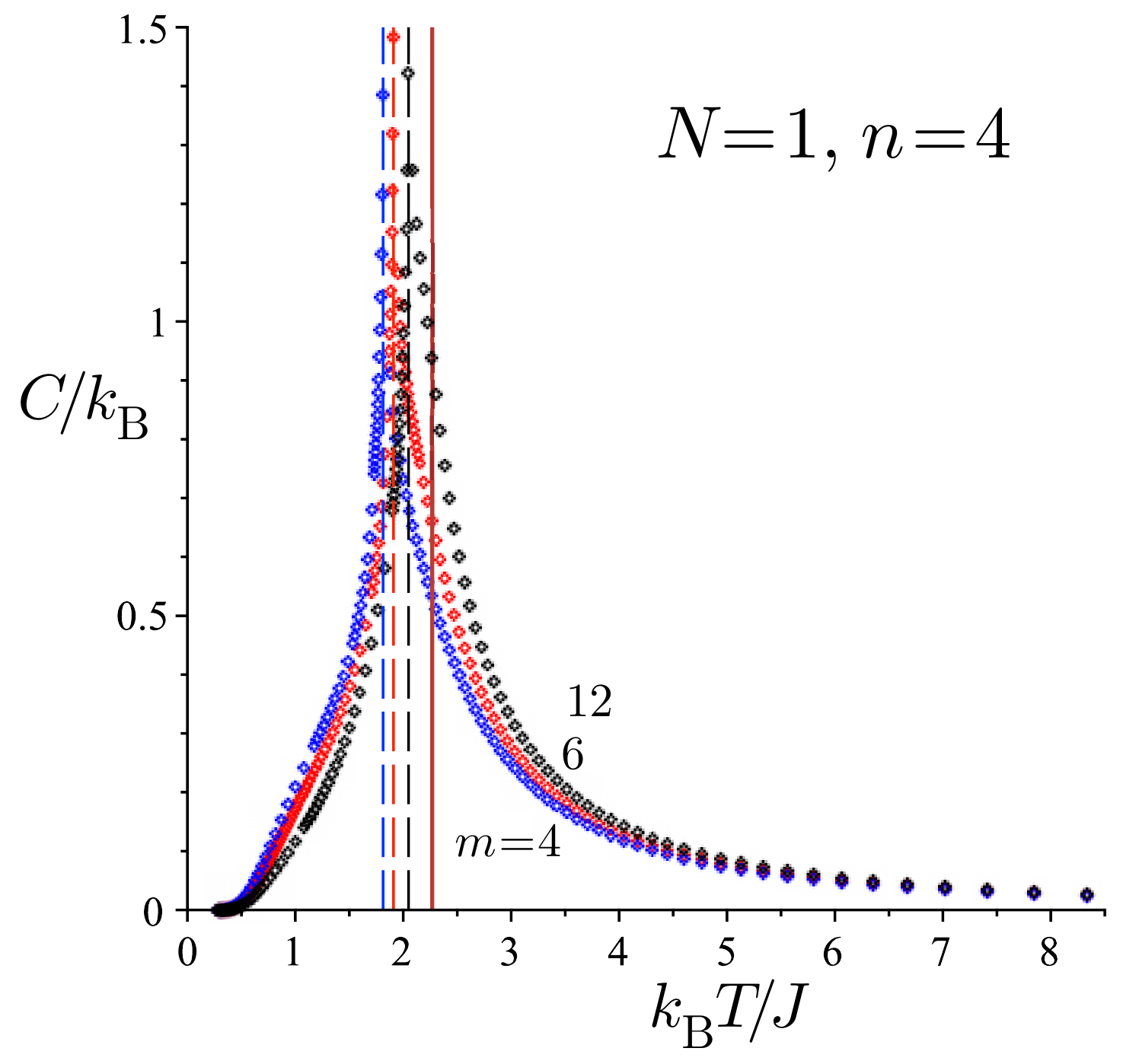}
        \includegraphics[width=0.48\hsize]{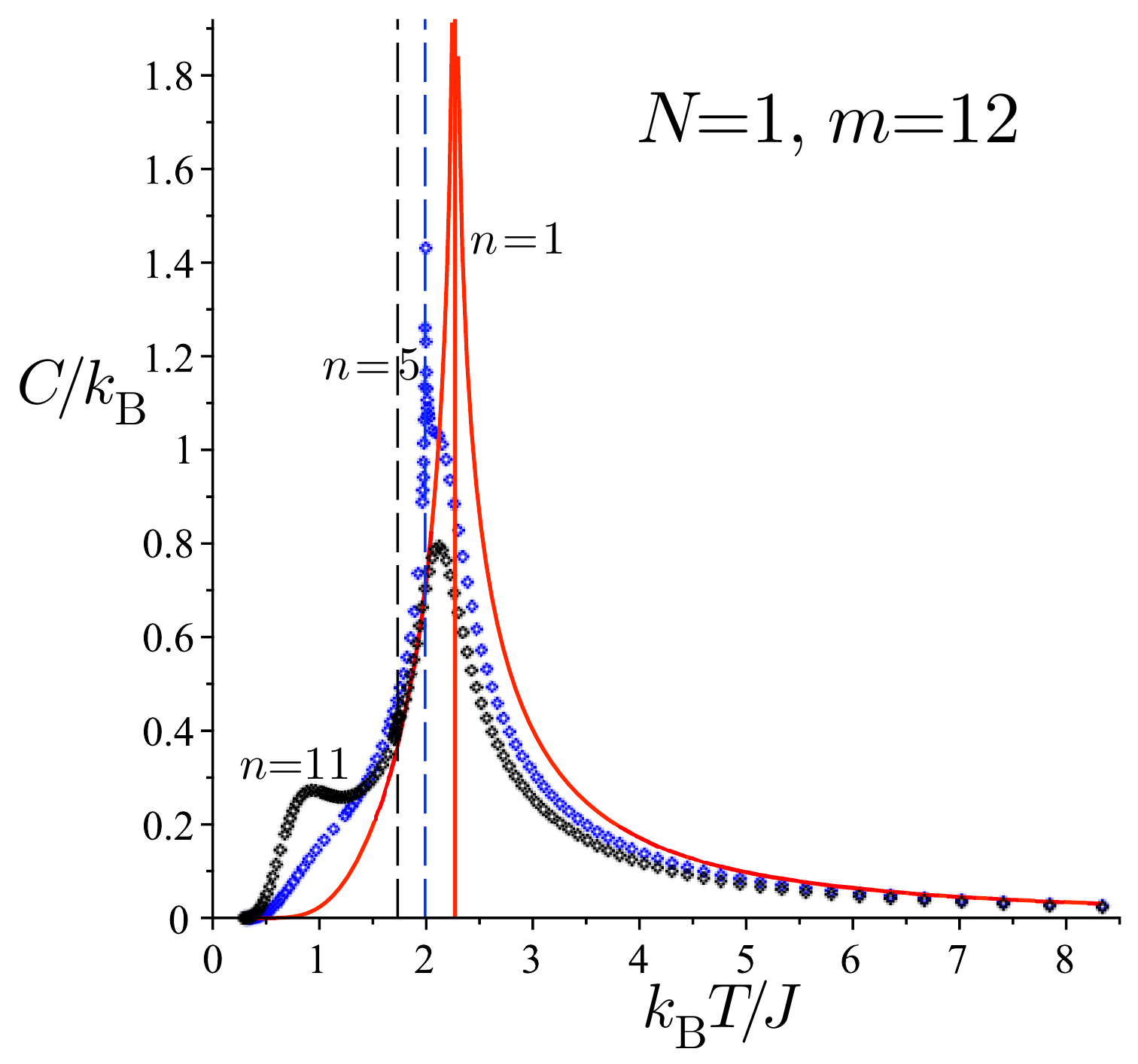}
\caption{(Color online) (a) The specific heats per site for separation of strings $N=1$ and length of strings $n=4$ are plotted for different strip widths $m=4,6,12$. We showed that the specific heat is logarithmically divergent at the bulk critical temperature $T_{\mathrm c}(1,m,4)$ denoted by dashed lines, which move toward the solid line (critical temperature of the Onsager Ising lattice) as $m$ increases. (b) The specific heats per site for $N=1$ and width of strip $m=12$ are plotted for $n=1,5,11$. In this case, the strings are next to each other. Thus for $n=1$, it is Onsager's Ising model, and its specific heat is plotted as a solid curved line. Its critical temperature is again represented by a solid vertical line as in (a). As $n$ increases, the bulk critical temperature becomes smaller. Only when $n>5$, the rounded peak due to finite strip width $m$ shows up. The amplitude of the logarithmic divergence at $T_{\mathrm c}(N,m,n)$, which dominates for $n\le 5$ shown in (a), decreases rapidly as $n$ increases. It becomes a small spike at the ``true" bulk critical point $T_{\mathrm c}(N,m,n)$ (indicated by the dashed vertical lines) at $n=5,11$. For $n=11$ a second peak due to the strings appears for $T\approx 0.8J/k_{\mathrm B}$.}
\label{fig:3}
\end{figure}

\subsection {N=2}
When the separation between the strings is $N=2$, the calculation is much more messy. For arbitrary $n$ the critical temperature $T_{\mathrm c}(N,m,n)$ for $N=2$ is lower than $T_{\mathrm c}(1,m,n)$, as shown in Fig.~2. It is determined from the following equation,
\be
 \tanh(x)\cosh[2(m+1)x]^{1/(m+n)}=1,\quad x=J/k_{\mathrm B}T.
\label{criticalT2} \ee
Particularly, because of missing bonds when $N=2$, the critical temperature $T_{\mathrm c}(2,m,n)$ for $n=1$ is lower than $T_{\mathrm c}(1,m,1)$, which is the Onsager Ising critical temperature for all $m$. Now we find that for $n\le3$ the specific heat diverges logarithmically without the rounded peak at lower temperature. The specific-heat behavior for $n=3$ and $m=4,6,12$ is shown in Fig.~4a.
 \begin{figure*}[htb]
\centering
\includegraphics[width=0.48\hsize]{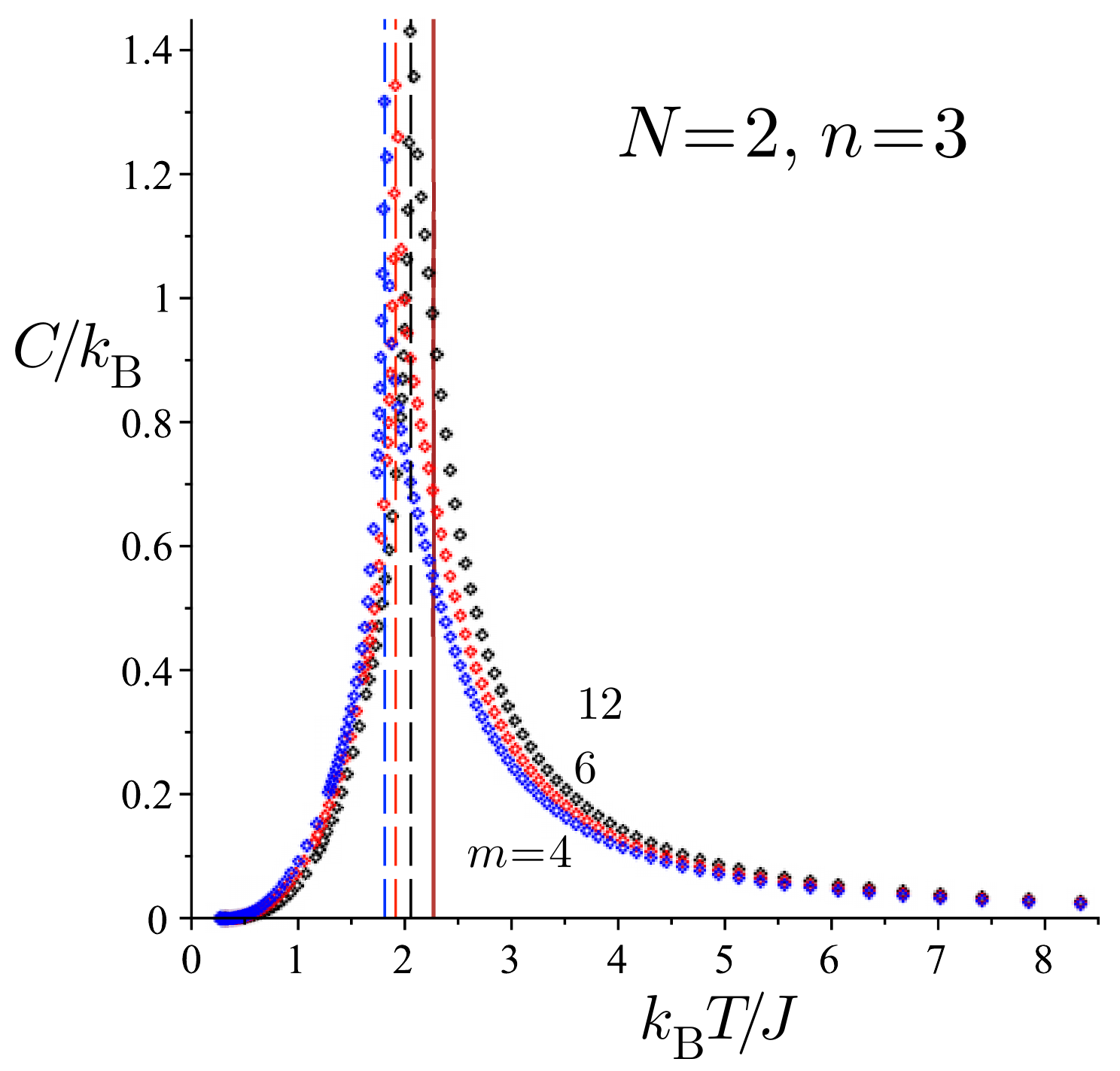}
        \includegraphics[width=0.48\hsize]{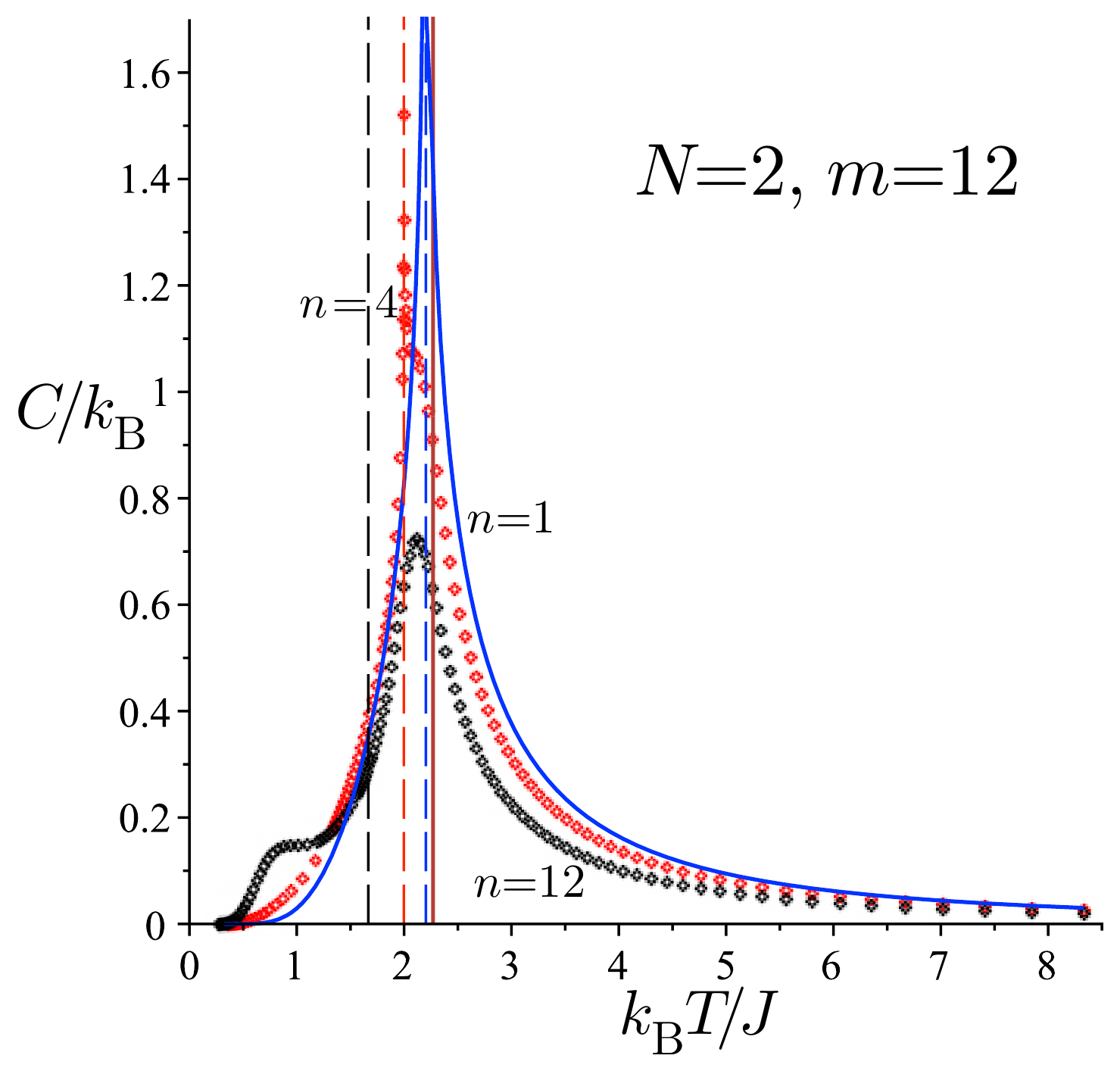}
\caption{(Color online) (a) Similar to the $N=1$ case shown in Fig.~3(a), we find that for $N=2$ and $n=3$ the specific heats per site diverge logarithmically without a rounded peak above the true bulk critical temperatures (denoted by dashed lines). These $T_{\mathrm c}(2,m,3)$ are lower than Onsager's critical temperature (solid vertical line), but approach it as $m$ increases.
(b) For $N=2$ and width $m=12$, the specific heats per site for $n=1$ diverge logarithmically at the bulk critical temperature $T_{\mathrm c}(2,12,n)$, (denoted by dashed lines); these are lower than Onsager's critical temperature (denoted by a solid line). As $n$ increases, the true bulk critical temperature $T_{\mathrm c}(2,12,n)$ becomes smaller and smaller. The specific heat diverges at $T_{\mathrm c}(2,12,n)$, but the amplitude of its logarithmic divergence becomes smaller and smaller as $n$ increases and is about invisible at $n=12$. The rounded peak due to the finite width $m=12$ shows up for both $n=4$ and $n=12$, while for $n=12$ another peak due to the strings appears below $T=J/k_{\mathrm B}$.}
\label{fig:4}
 \end{figure*}
In Fig.~4(b), we again plot the specific heat, but now for fixed $m=12$ and different $n$. Similarly as for $N=1$, we find the rounded peak in the specific heat to show up for $n\ge4$, while the logarithmic divergence at the true critical temperature $T_{\mathrm c}(2,m,n)$ has a diminishing amplitude as $n$ increases. This again shows one-dimensional behavior of the finite-width strips, which becomes visible as $n$ increases. A second peak at lower temperatures appears when the length $n$ of the strings becomes large, see the plot for $n=12$.
Even though we find no strong difference between the cases of $N=1$ and $N=2$, it is much more difficult to calculate the spontaneous magnetization for $N=2$. However, it easily seen that it is nonzero for $T<T_{\mathrm c}(2,m,n)$. Since it is a two-dimensional Ising model, we can use universality arguments to argue that the critical exponents are the same as for the Onsager Ising lattice. Particularly, we conclude that the spontaneous magnetization $M$ approaches 0 as $T\to T_{\mathrm c}(2,m,n)$ with the same $\beta=1/8$ power law. 
\subsection {N=3}
The calculations for $N=3$ are very complicated and messy. In this case, the critical temperatures are determined from equations which are much more difficult to derive than for $N=1,2$, and they are 
\ba
F(z)\bigg[1-\frac{z^{m+n}}3\Big(\frac{1+z}{1-z}\Big)^{m+1}\bigg]=\frac{2 z^n}3,
 \ea
 where
 \ba
 F(z)= \frac{(1+z^2+z)}{(1-z^2)}\bigg[\frac{\alpha_1^{m+1}-\alpha_1^{-m-1}}{\alpha_1-\alpha_1^{-1}}\bigg] -z\bigg[\frac{\alpha_1^{m}-\alpha_1^{-m}}{\alpha_1-\alpha_1^{-1}}\bigg],
 \ea
 with
 \ba
 \alpha_1=r+\sqrt{r^2-1},\qquad r=\frac{(z^2+1)^2}{2z(1-z^2)}+\half.
 \ea
To calculate the free energy is even more difficult than the critical temperature. Even when the integrals for the specific heats are obtained, to plot the results requires a lot of computing time and we needed to split the calculations to small pieces for our computers to handle. But the behaviors found are very similar to those of the $N=1$ and $N=2$ cases. Namely, the specific heat has logarithmic divergence at $T_{\mathrm c}(3,m,n)$ without a rounded peak for $n\le2$, and the rounded peak shows up as $n$ increases. In Fig.~5(a), we plotted the specific heats per site for fixed string length $n=6$, but different widths $m$ of the strips. We find, as $m$ increases, that the rounded peak becomes more prominent and moves toward the Ising critical temperature, while the amplitude of the logarithmic divergence at the true bulk critical temperature becomes smaller as $m$ increases.
\begin{figure*}[htb]
\centering
\includegraphics[width=0.48\hsize]{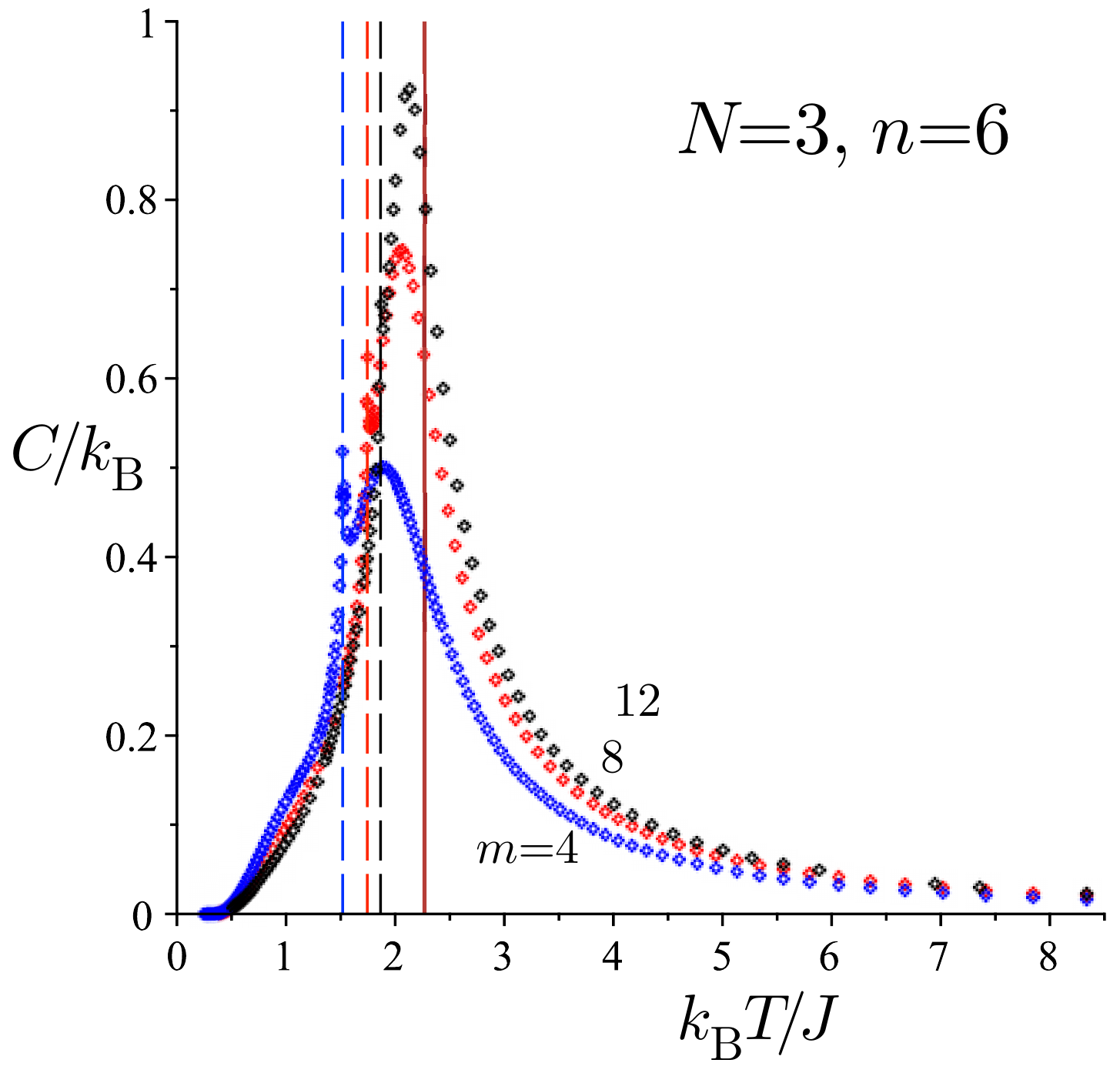}
        \includegraphics[width=0.48\hsize]{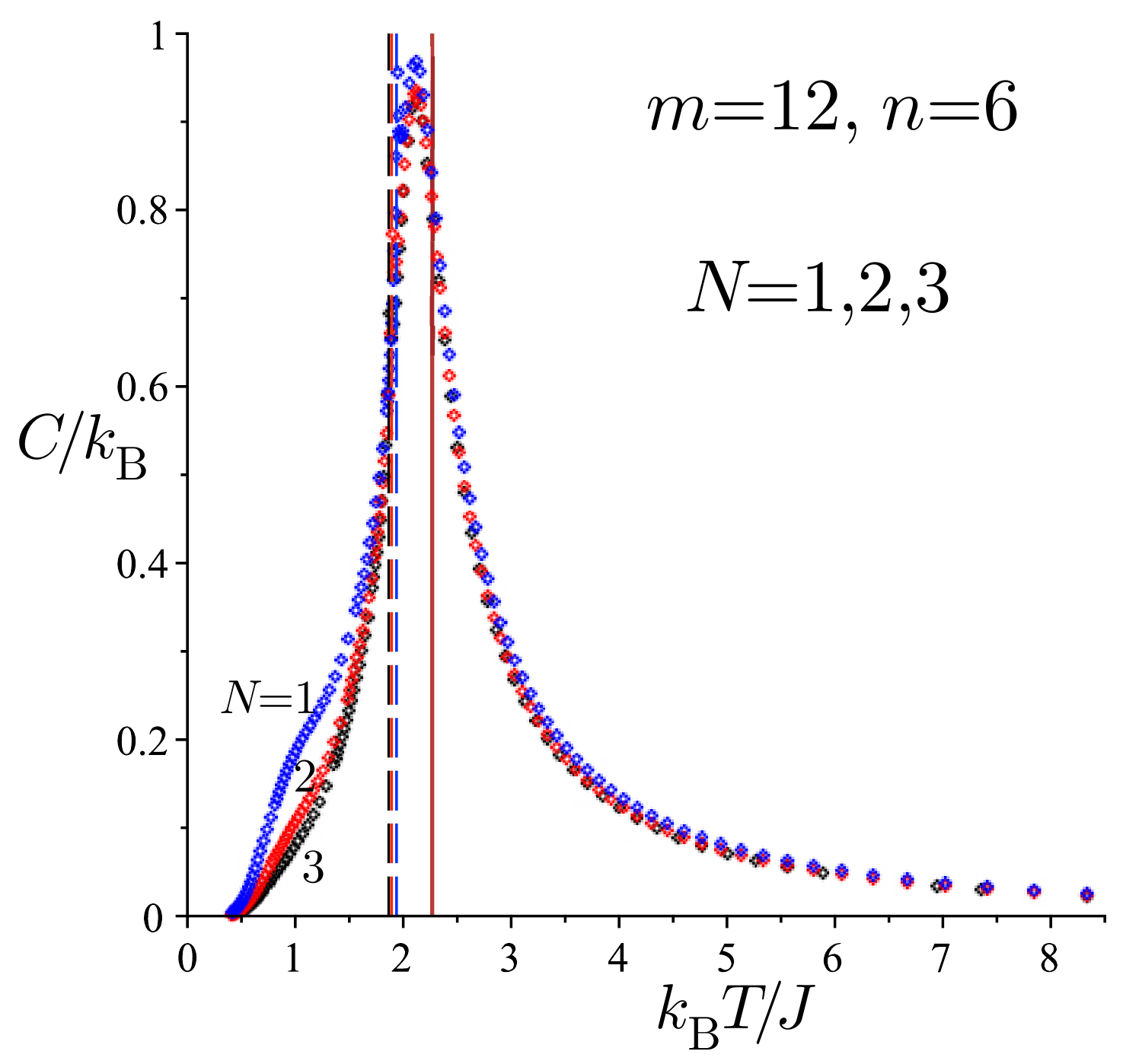}
\caption{(Color online) (a) We plot the specific heat as a function of temperature for fixed $N=3$, $n=6$, but different $m$. We again denote the true bulk critical temperatures $T_{\mathrm c}(3,m,6)$, which increase as $m$ increases, by dashed vertical lines. We find that the amplitude of the logarithmic divergence becomes small as $m$ increases, and the rounded peak moves toward the solid vertical line, which corresponds to the critical temperature of Onsager's Ising model. (b) We also plotted the specific heat versus temperature for fixed $m=12$ and $n=6$, but for different $N$. For $N=1$ which is represented by the blue dots, there is a visible spike for the specific heat at its bulk critical temperature $T_{\mathrm c}(1,12,6)$; for $N=2,3$, however, the logarithmic divergences are about invisible.}
\label{fig:5}
 \end{figure*}

The calculations for $N\ge3$ are very complicated and messy, but may not provide more insight to
physics. To see the difference due to the separation $N$ between strings, we plot in Fig.~5(b) the specific heat versus temperature for fixed $m=12$ and $n=6$, but for different $N$. The bulk critical temperature does not change much and we find $T_{\mathrm c}(1,12,6)=1.938063784$, at which there is a visible spike for specific heat; for $N=2,3$, we have $T_{\mathrm c}(2,12,6)=1.891784286$ and $T_{\mathrm c}(3,12,6)=1.865375064$, at which the logarithmic divergences are scarcely visible. Also, apart from a small shoulder at lower temperature for $N=1$, the plots fall almost on top of one another.
This is the most surprising result. This shows the separations $N$ between the strings are less important than the lengths $n$ of the strings. For $N=1$, the correlation at the center of the strip can be calculated. As that would reveal a great deal more about the proximity effect, we address this next.
\section{Spontaneous Magnetization and Spin Correlations in a Central Row}
The previous calculation shows that the case with separation of strings $N=1$ gives as much information on the behavior of the system as those with larger separations between the strings. For $N=1$, the row correlation of spins at the center of a strip of width $m=2j$\footnote{To have a row at the center, $m$ needs to be even. Then the model is reflection invariant about this row and translation invariant in the horizontal direction, so that (\ref{Toeplitz}) follows.} is given as the Toeplitz determinant,\footnote{For some early works on the magnetization in layered Ising models see \cite{WZ,KR}.}
\ba
\langle \sigma_{0,1}\sigma_{0,r+1}\rangle=\left|\begin{array}{ccccc}
a_0&a_{-1}&a_{-2}&\cdots&a_{1-r}\\
a_1&a_0&a_{-1}&\cdots&a_{2-r}\\
a_2&a_1&a_0&\cdots&a_{3-r}\\
\vdots&\vdots&\vdots&\ddots&\vdots\\
a_{r-1}&a_{r-2}&a_{r-3}&\cdots&a_{0}
\end{array}\right|,
\label{Toeplitz}
\ea
where 
\ba 
a_n=\frac 1{2\pi}\int_{-\pi}^{\pi}\rd \theta\, \re^{-i n\theta}\,\Phi(\theta),
\quad \Phi(\theta)=\sqrt\frac{\overline {A(\theta)}\,\overline {B(\theta)}}{A(\theta)B(\theta)},
\label{Phi}\ea
in which ${\bar f}$ denotes the complex conjugate of $f$, and
\ba
A(\theta)=\rho_a\prod_{\ell=1}^{j+1}(1-{\hat\gamma}_\ell\re^{-i\theta} )
               \prod_{\ell=j+2}^{2j+1}(1-{\hat\gamma}^{-1}_\ell\re^{i\theta} ),\cr
 B(\theta)=\rho_b\prod_{\ell=1}^{j}(1-{\gamma}_\ell\re^{-i\theta} )
               \prod_{\ell=j+1}^{2j+1}(1-{\gamma}^{-1}_\ell\re^{i\theta} ).
\label{AB}\ea

To have a central row, we need to have $m$ even, that is $m=2j$.
Unlike the row correlation of the Onsager lattice, where the generating function has only two roots
$\gamma_1$ and $\hat\gamma_1$, which can be explicitly calculated, the $2j+1$ roots in (\ref{AB}) of these Laurent polynomials in $e^{i\theta}$ can only be calculated numerically. From these calculations, we find that all the roots are real, $A(\theta)$ has $j+1$ roots smaller than 1, and $j$ roots greater than 1 for all temperatures, while $B(\theta)$ has $j+1$ roots smaller than 1 and $j$ roots greater than 1 for $T>T_{\mathrm c}(1,m,n)$, but one of the roots, say $\gamma_{j+1}$, becomes 1 at the critical temperature, and greater than 1 for $T<T_{\mathrm c}(1,m,n)$. In (\ref{AB}), we let $j+2\le\ell\le 2j+1$ be the subscript to denote the $j$ roots which are always greater than 1.

Even though these formulae look formidable, it is possible to calculate the spontaneous magnetization using Szeg\H o's theorem. We find that the spontaneous magnetization ${ M}$ at the center of the strip is of the form
\ba
{M}=(1-\gamma^{-2}_{j+1})^{1/8}{\cal G}_m
\ea
where ${\cal G}_m$ is a complicated expression involving the $2(m+1)$ roots of $A(\theta)$ and $B(\theta)$, which will be given in a later paper.
We plotted in Fig.~6(a) this spontaneous magnetization at the central row of the strip for fixed string length $n=7$, but for strips of different widths $m=4,8,12$. Even though, as $m$ increases, we need to calculate more roots, which requires more digits for accuracy, we find the magnetization starts from 1, and drops to zero sharply with the 2-d exponent $\beta=1/8$ as $T$ approached its respective critical temperature $T_{\mathrm c}(1,m,n)$.
In Fig.~6(b), the magnetization is plotted for fixed $m$, but for different $n$, demonstrating the same behavior.
 \begin{figure*}[htb]
\centering
\includegraphics[width=0.48\hsize]{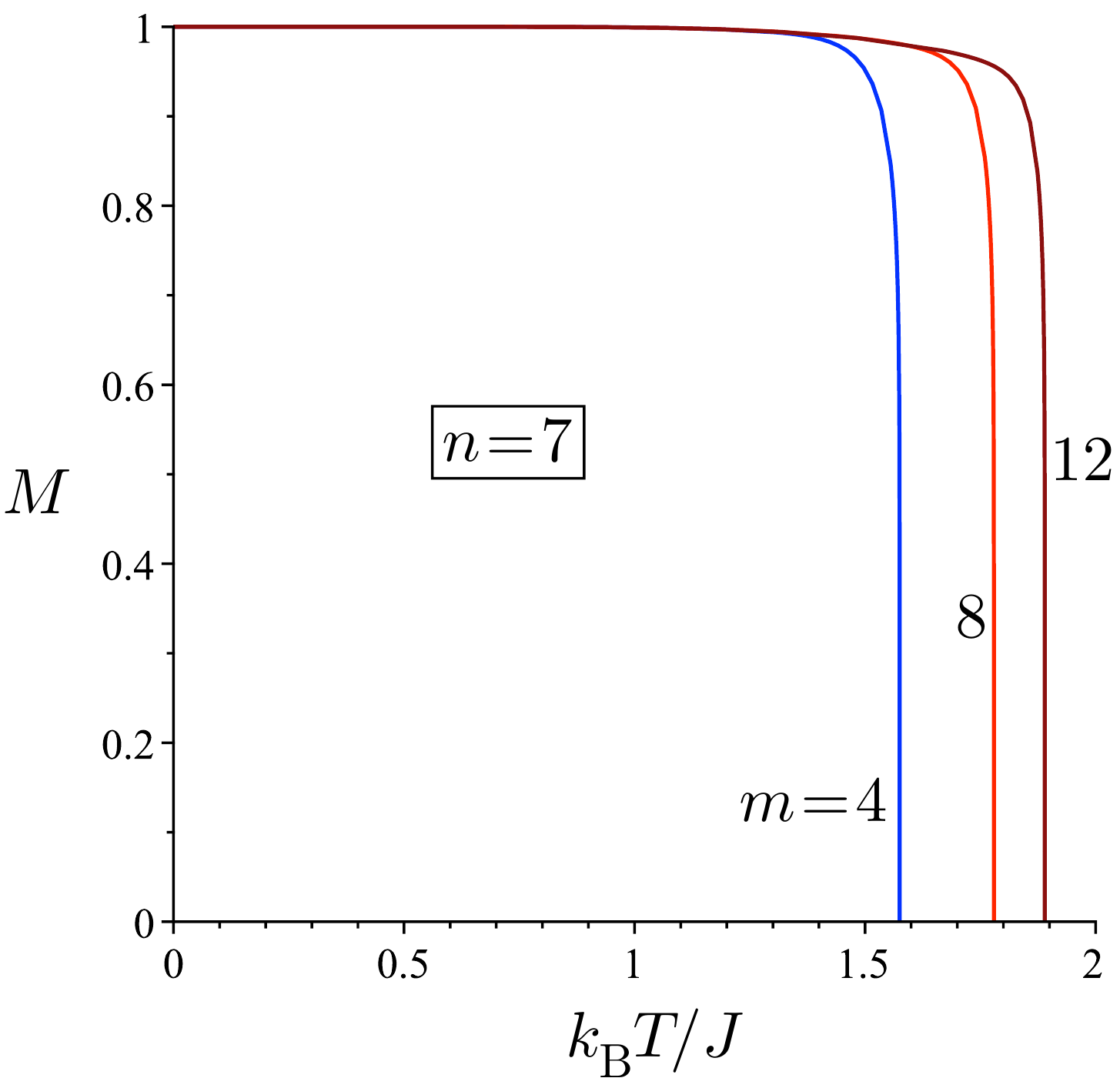}\hspace{10pt}
        \includegraphics[width=0.48\hsize]{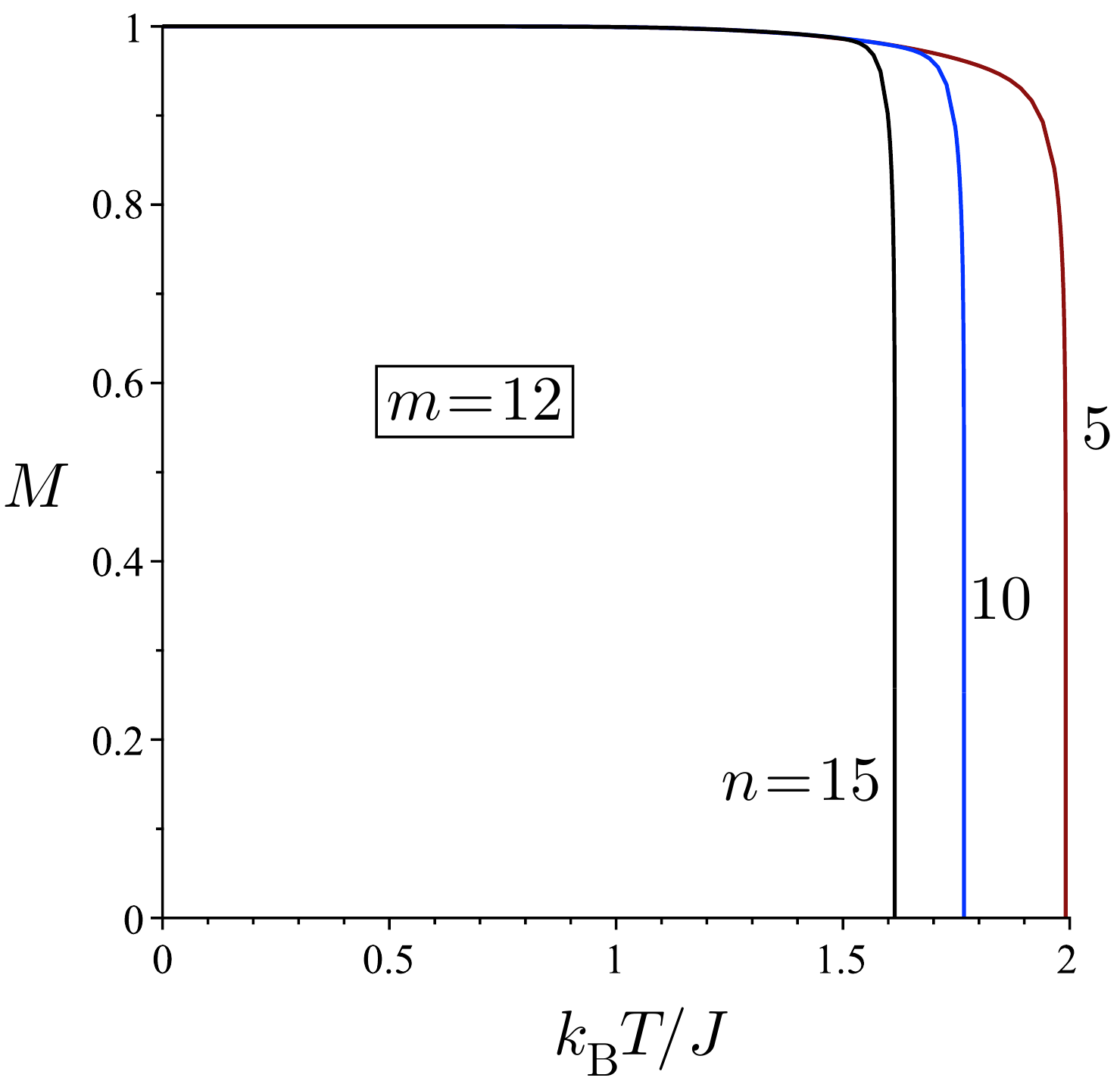}
\caption{(Color online) (a) The spontaneous magnetization $M$ in the center row of a strip is plotted as a function of temperature, for string length $n=7$ and strip width $m=4,8,12$. 
(b) For fixed $m=12$, we plotted the spontaneous magnetization as a function of temperature $T$ for $n=5,10,15$.}
\label{fig:6}
 \end{figure*}

We can also calculate the asymptotic behavior of the spin-spin correlation in the center row of a strip for separation $r$ between the spins large and for $T$ near $T_{\mathrm c}(1,m,n)$.\footnote{The separation must be large compared to the correlation length in the scaling region near $T_{\rm c}$.}\
For $T>T_{\mathrm c}(1,m,n)$, we have $\gamma_{j+1}<1$, and $\gamma_{j+1}>1$ for $T<T_{\mathrm c}(1,m,n)$, we find
\ba
\langle \sigma_{0,1}\sigma_{0,r+1}\rangle=\frac{\gamma^N_{j+1}{\cal H}_m^+}{ \sqrt r}+\cdots,\hspace{30pt}\quad 
T>T_{\mathrm c}(1,m,n),
\label{C-above}\\
\langle \sigma_{0,1}\sigma_{0,r+1}\rangle={M}^2\bigg[1+\frac{\gamma_{j+1}^{-2N}{\cal H}_m^-}{r^2}+\cdots\bigg],\quad 
T<T_{\mathrm c}(1,m,n),
\label{C-Below}\ea
which is identical to the expressions (2.43) on page 243 and (3.23) on page 260 of the book by McCoy and Wu \cite{MWbk}, except that the functions ${\cal H}_m^{\pm}$ are now complicated functions of the $2(m+1)$ roots of $A(\theta)$ and $B(\theta)$. This demonstrates the same exponents $1/2$ and 2 as those of the regular 2-d Ising model.

As we have $\gamma_{j+1}<1$ for $T$ greater than the true critical temperature $T_{\mathrm c}(1,m,n)$, we find from (\ref{C-above}) that the inverse correlation length $\xi^{-1}=\ln \gamma^{-1}_{j+1}$, so that the correlation decays as $\re^{-r/\xi}$. On the other hand, below the true critical temperature, we have $\gamma_{j+1}>1$. Then from (\ref{C-Below}) the true correlation length is $\xi/2$ with $\xi^{-1}=\ln \gamma_{j+1}$, so that the correlation now decays as $\re^{-2r/\xi}$ \cite{Kadanoff,Wu}. 

It is easily seen from (\ref{criticalT1}), that for systems with the same ratio $(n-1)/(m+1)$ have the same critical temperature. Thus by comparing these systems, we may gain insight to dependence of the correlation length as a function of $m$. Particularly for $n=m+2$ (e.g.\ $m=4$ and $n=6$, or $m=12$ and $n=14$), the ratio is 1 and the critical temperature is $T_{\mathrm c}(1,m,m+2)=T_{\mathrm c}(1,12,14)=1.641017930$. For the deviation from criticality we shall use the often used $t=1-T/T_{\mathrm c}$ or $T=T_{\mathrm c}(1-t)$, so that $t>0$ when $T<T_{\mathrm c}$.

In Fig.~7 we plot the inverse correlation lengths $\ln\gamma^{\pm}_{j+1}$ as a functions of $|t|=|1-T/1.641017930|$, for $m=4 (n=6)$, $m=8 (n=10)$ and $m=12 (n=14)$.
As $m$ increases, we find that the correlation lengths become larger, which is represented by the lower curves in Fig.7a.\footnote{At some larger values of $t$ the curves for $T<T_{\mathrm c}$ cross those for $T<T_{\mathrm c}$, since $t=\pm1$ represent $T=0$ (where $\xi=0$) and $2T_{\mathrm c}$ (where $\xi$ is still finite).}

\begin{figure}[htb] 
\centering
  \vspace*{0pt}
     \includegraphics[width=0.47\hsize]{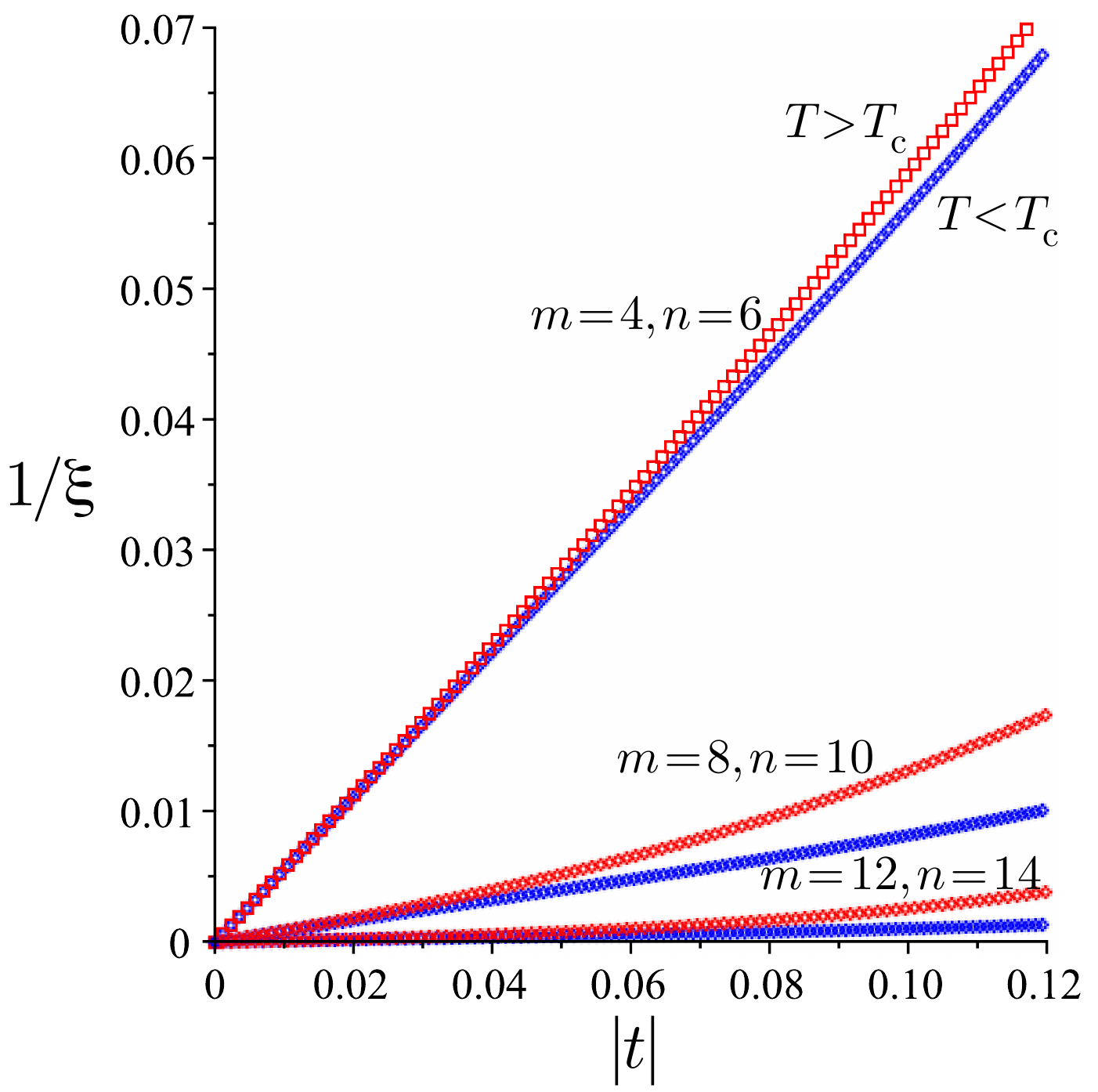}
        \includegraphics[width=0.47\hsize]{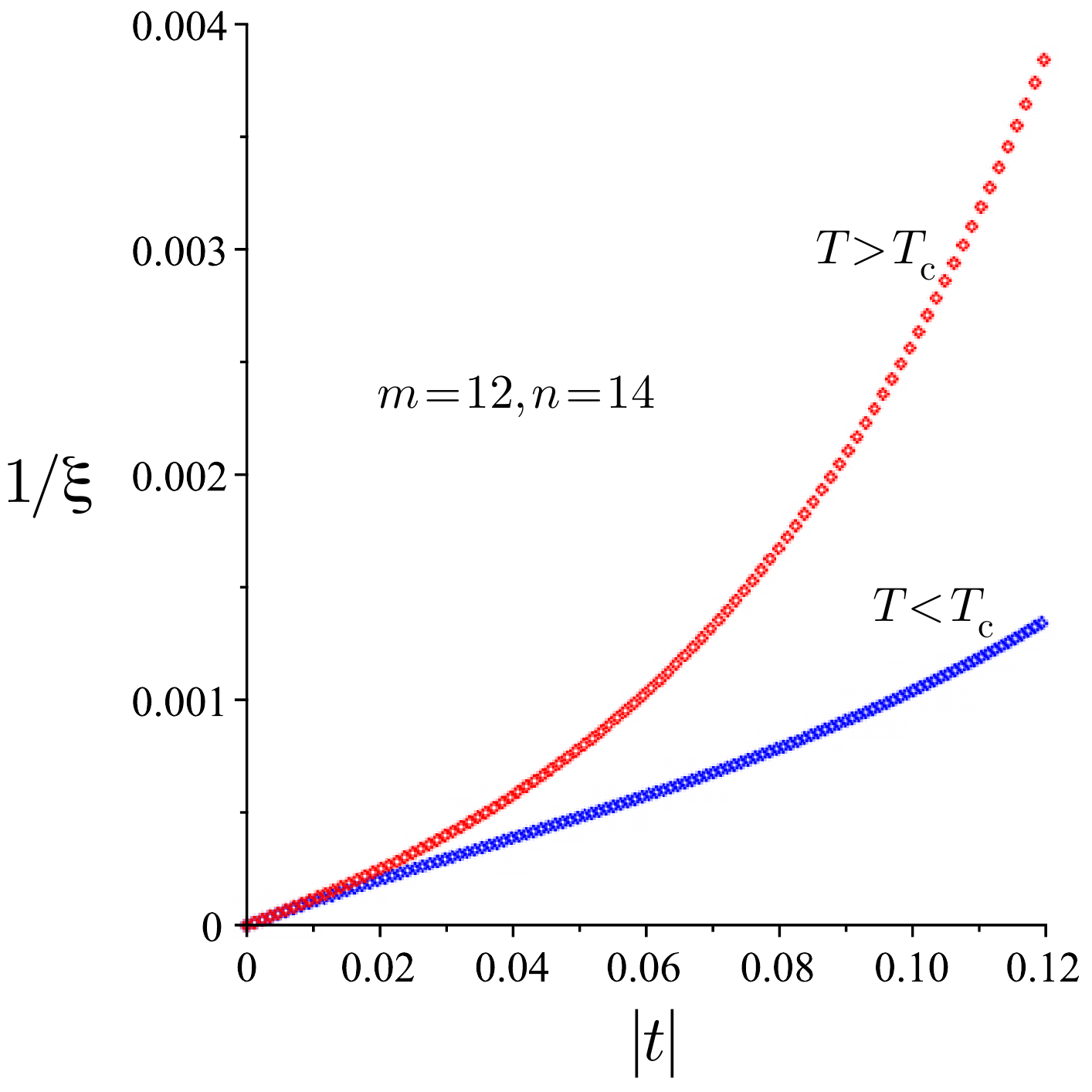}
\caption{(Color online) (a) The inverse correlation lengths $1/\xi$ are plotted for $m=4(n=6)$, $m=8(n=10)$ and $m=12(n=14)$; The red points are $1/\xi=\ln \gamma^{-1}_{j+1}$ for $T>1.641017930$, and the blue points are
$1/\xi=\ln \gamma_{j+1}$.
(b) Enlarged figure for $m=12(n=14)$, which shows that, as $m$ increases, the regime $1/\xi\propto |t|$, shrinks.}
\label{fig:7}
\end{figure}
To understand the corrections to scaling near critical temperature $T_{\mathrm c}(1,m,m+2)=1.641017930$ we expand the inverse correlation lengths for $n=m+2$ in Taylor series,
\ba\fl
\frac1{\xi}=\ln \gamma^{\mp1}_{j+1}=\begin{cases} {
  0.5590194|t|\pm0.1761049|t|^2+1.7413970|t|^3,&\hbox{$m=4$},\\
  0.0877259|t|\pm0.2328992|t|^2+1.8826931|t|^3,&\hbox{$m=8$},\\
  0.0110719|t|\pm0.0563571|t|^2+0.6323108|t|^3, &\hbox{$m=12$},}
                                                                    \end{cases}
                                                                    \ea
with sign choices corresponding to $T\gtrless T_{\mathrm c}(1,m,m+2)$.
This shows that as $m$ increases and $n=m+2$, the correlation length of the spins on the central row increases. The second and third terms in these
expansions are corrections to scaling, whose coefficients become larger than the one of the leading term. To illustrate this more clearly, we have enlarged the plot for $m=12$ and $n=14$ in Fig.~7b.

We shall next include some mathematical details to show the dependence of the generating function on $m$ and $n$ in order to demonstrate the possibility of calculating the scaling function.
\subsection{Limiting cases}
In calculating the correlation function, we chose to make the vertical and horizontal couplings to be different in order to distinguish the vertical and horizontal correlation lengths. We denote the horizontal coupling by $J'$, and $z'=\tanh J'/k_{\mathrm B}T$. We also need the variable
\be z^*=(1-z)/(1+z)=\re^{-2J/k_{\mathrm B}T},
\label{dual}\ee
related to the dual variable of the Kramers--Wannier duality transform.

The functions in (\ref{Phi}) are given by
\ba\fl
A(\theta)\!=\!(\alpha^j+\alpha^{-j})[z'(z^n-1)\re^{-i\theta}+(z^n+1)]+\Omega^{-\halfs}(\alpha^j-\alpha^{-j})[(z^n-1)\re^{-i\theta}+z'(z^n+1)],\cr
\fl
B(\theta)\!=\!(\alpha^j+\alpha^{-j})[(z^n-1)\re^{i\theta}+z'(z^n+1)]+\Omega^{-\halfs}(\alpha^j-\alpha^{-j})[z'(z^n-1)\re^{i\theta}+(z^n+1)],\cr
\label{ABp}\ea
where\footnote{This $\alpha$ is related to the integrand of the free energy of the perfect Ising model, and it is the $\alpha_i $ of \cite{HAY}, where it is expressed in terms of $t\propto (T/T_{\mathrm c}-1)$. However, for the correlation function it has to be written in a different form.} 
\ba
\alpha^{\pm 1}=G\pm\sqrt{G^2-1},\cr
 G= [(1+{z'}^2)(1+{z^*}^2)-4z'z^*\cos\theta]\Big/[(1-{z'}^2)(1-{z^*}^2)].
\label{G} \ea
It can be easily verified that
 \ba
 {G^2-1}&=&\frac{4(1-z'z^*\re^{i\theta})(1-z'z^*\re^{-i\theta})(z'-z^*\re^{i\theta})
 (z'-z^*\re^{-i\theta})}{(1-{z'}^2)(1-{z^*}^2)}\label{G}\\
 &=&\frac{4(1-z'z^*\re^{-i\theta})^2(z'-z^*\re^{i\theta})^2\Omega}{(1-{z'}^2)(1-{z^*}^2)},
 \label{GOmega}\ea
 so that
 \be \Omega=\frac{(1-z'z^*\re^{i\theta})(z'-z^*\re^{-i\theta})}{(1-z'z^*\re^{-i\theta})(z'-z^*\re^{i\theta})},\qquad \Omega^{-1}=\overline \Omega.
\label{Omega}\ee
In (\ref{G}) and (\ref{Omega}), we used the same form as used by Baxter in his most recent paper \cite{BaxIS}. 

In the limit $m=2j\to\infty$, we may drop $\alpha^{-j}$ in (\ref{ABp}), and find
\be
\overline{A(\theta)}=\Omega^{\halfs}B(\theta),\quad
 \overline{B(\theta)}=\Omega^{\halfs}A(\theta),
 \ee
 so that the generating function in (\ref{Phi}) becomes
 \be \Phi(\theta)=\Omega^{\halfs},\label{PhiMW}\ee
which is identical to the generating function given in (1.3) and (1.4) on page 249 in McCoy and Wu's book \cite{MWbk} for the row correlation, as it should.
In this limit $m\to\infty$, the infinitely wide strip has its horizontal and vertical correlation length given by\footnote{We ignore here the anomaly that below $T_{\mathrm c}$ the correlation length should have an extra factor $\frac12$ \cite{MWbk}.}
\be \fl\frac1{\xi^{\pm}_{\mathrm h}}=\pm\ln \Bigg[{z'}\,\frac{1+z}{1-z}\Bigg]=\pm\ln \Bigg[\frac{z'}{z^*}\Bigg], \quad
\frac1{\xi^{\pm}_{\mathrm v}}=\pm\ln \Bigg[{ z}\,\frac{1+z'}{1-z'}\Bigg]=\pm\ln \Bigg[\frac{z}{{z'}^*}\Bigg],\label{corrlengths}\ee
with $+$ for $T>T_{\mathrm c}$ and $-$ for $T<T_{\mathrm c}$.
 
In the limit $n\to\infty$, we have $z^n\to0$, and (\ref{ABp}) becomes
 \ba
A(\theta)=(\alpha^j+\alpha^{-j})(1-z'\re^{-i\theta})+\Omega^{-\halfs}(\alpha^j-\alpha^{-j})(z'-\re^{-i\theta}),\cr
B(\theta)=(\alpha^j+\alpha^{-j})(z'-\re^{i\theta})+\Omega^{-\halfs}(\alpha^j-\alpha^{-j})(1-z'\re^{i\theta}).\label{ABz}\ea
Therefore,
 \be
 B(\theta)\to-\re^{i\theta}A(\theta),\qquad\overline{B(\theta)}\to-\re^{-i\theta}\overline{A(\theta)},
 \ee
 so that
 \be
 \Phi(\theta)=\sqrt{\re^{-2i\theta}\frac{\overline{A(\theta)}^2}{A(\theta)^2}}=-\re^{-i\theta}\frac{\overline{A(\theta)}}{A(\theta)}.
 \label{Phiz0}\ee
 The choice of sign is to make $-\re^{-i\pi}=1$. Because the square root disappears, its correlation function behaves very differently from (\ref{C-above}), decaying exponentially as in the one-dimensional Ising model. 
 
The full spin-spin correlation of the single finite-width strip case, resulting from this limit $n\to\infty$, is not known. Obviously, it differs from row to row. In fact, it is known that, except for the center-row case above, it may be given in terms of block-Toeplitz determinants \cite{HAYMcCoy2}. However, taking a second limit $m=2j\to\infty$, we can drop $\alpha^{-j}$ as before, and find from (\ref{ABz}) that
\be
A(\theta)=\alpha^j[(1-z'\re^{-i\theta})+\Omega^{-\halfs}(z'-\re^{-i\theta})],\quad
\overline{A(\theta)}=-\Omega^{\halfs}\re^{i\theta}A(\theta).
\ee
Consequently, the generating function in (\ref{Phiz0}) becomes (\ref{PhiMW}), reproducing the 2-d behavior again as it should. More generally there is a crossover for finite $m$: If $\alpha$ is expressed in terms of $t\propto T/T_{\mathrm c}-1$ as in \cite{HAY}, then $\alpha^{-j}$ is exponentially small when $m|t|$ is large, and the system behaves as two-dimensional, otherwise it acts one-dimensional. This shows that it is possible to study the behavior of the correlation function as a function of the scaling variable $|t|m$. However, to calculate the correlation function, we need to make a Wiener--Hopf splitting of the generating function, which may be very difficult when $\alpha$ is expressed in the scaling form.
\section{Open Problems : }
\subsection{Correlation Function of a Single Strip of Finite Width}
The correlation for the central row of a single strip of width $m$ is a Toeplitz determinant whose generating function is given in (\ref{Phiz0}), but the correlations within other rows are different, and may be expressed as block-Toeplitz determinants. How to calculate these block-Toeplitz determinants is a very challenging problem.
Furthermore, even for the central row, the generating functions are ratios of two polynomials with degree $m+1$. As $m$ increases, one needs to calculate more and more roots. We have shown that in the limit $m\to\infty$, the generating function becomes the well-known square-root function (\ref{PhiMW}) for the correlation function of Onsager's 2-d Ising model. Therefore, it would be very difficult, but most interesting, to study the scaling behavior of these correlations in the limit, $m\to\infty$ and $T\to T_{\mathrm c}$, where $T_{\mathrm c}$ is Onsager's critical temperature given by (\ref{OnsagerTc}). One still expects that for $m|T/T_{\mathrm c}-1|\ll1$, the correlation function behaves like that of a one-dimensional system, but behaves as that of a two-dimensional system in the opposite limit. To express the correlation in the scaling regime was already highly nontrivial for the original Ising model \cite{WMTB} with $m=\infty$. 

 \subsection{Scaling Functions for Strips connected by Strings}
When we consider an infinite system of horizontal strips of width $m$ connected by sequences of strings of finite length $n$ as in Fig.~1, the behavior changes a great deal. For $n\le 4$, we found that the specific heat diverges at $T_{\mathrm c}(1,m,n)$ logarithmically for all values of $m$. However, as $n$ increases, rounded peaks in the specific heat appear above this temperature signifying the one-dimensional behavior of the strips. The spontaneous magnetization is nonzero for $T<T_{\mathrm c}(1,m,n)$. These results show that the scaling functions for the specific heat and correlations are much more complicated---namely in addition to the dependence on the scaling variable $m/\xi_h=m|T/T_{\mathrm c}-1|$, the dependence on another scaling variable related to the length $n$ of the strings must be added.

For the vertical one-dimensional strings, the critical temperature is at $T=0$ and it is well-known that their inverse correlation length $1/\xi^+_{\mathrm s}=\ln z$. The critical temperature equation in (\ref{criticalT1}) for our rectangular Ising model with holes with different horizontal and vertical couplings generalizes to
\be
z^{n+m}\,\Bigg[\frac{1+z'}{1-z'}\Bigg]^{m+1}=1
\ee
Taking the log of this and using (\ref{corrlengths}) this can be rewritten as
\be
\frac{n-1}{\xi^+_{\mathrm s}}-\frac{m+1}{\xi_{\mathrm v}^{-}}=0,
\label{xiTc}\ee
suggesting the possible additional scaling variable to be $(n-1)/\xi_s^+$. This seems to agree with the observation given in \cite{AMSV}. Another problems concerns the distribution of the roots for $A(\theta)$ and $B(\theta)$. The statement that these two functions have $j+1$ roots smaller than 1 and $j$ roots greater than 1 for $T>T_{\mathrm c}(1,m,n)$ is based on numerical evidence. There remain also many other challenging and difficult problems such as the spin-spin correlation at the other rows and the magnetic susceptibility of the system.

\end{document}